\documentclass[aps,prd,showpacs,amsmath,amssymb,nofootinbib,eqsecnum, preprintnumbers,twocolumn]{revtex4-1}
 
\usepackage{hyperref} 
\usepackage{graphicx,xcolor}
\usepackage{epstopdf}

 \newcommand{\sgn}{\mathop{\mathrm{sgn}}}

\newcommand{\be}{\begin{equation}}
\newcommand{\ee}{\end{equation}}
\newcommand{\ba}{\begin{eqnarray}}
\newcommand{\ea}{\end{eqnarray}}
\newcommand{\nn}{\nonumber} 
\newcommand{\cH}{\mathcal{H}} 
 
\newcommand{\cM}{\mathcal{M}} 
\newcommand{\dl}[1]{\frac{d #1}{d \lambda}} 
\newcommand{\dlb}[1]{\frac{d #1}{d \bb{\lambda}}}

\newcommand{\h}[1]{\hat{#1}} 
\newcommand{\bb}[1]{\bar{#1}}

 \newcommand{\dm}{n}

\newcommand{\ptm}{\mathbb{P}T\cM}

\begin{document}
\title{Null lifts and projective dynamics}

\author{Marco Cariglia}
\email{marco@iceb.ufop.br}
\affiliation{Universidade Federal de Ouro Preto, ICEB, Departamento de F\'isica.
  Campus Morro do Cruzeiro, Morro do Cruzeiro, 35400-000 - Ouro Preto, MG - Brasil}

\date{\today}  

\begin{abstract} 
We describe natural Hamiltonian systems using projective geometry. The null lift procedure endows the tangent bundle with a projective structure where the null Hamiltonian is identified with a projective conic and induces a Weyl geometry. Projective transformations generate a set of known and new dualities between Hamiltonian systems, as for example the phenomenon of coupling-constant metamorphosis. We conclude outlining how this construction can be extended to the quantum case for Eisenhart-Duval lifts. 
\end{abstract}

\maketitle

\section{Introduction} 
Hamiltonian dynamics is a classic subject that, despite its age, does not cease to display new phenomena and provide insights. One relatively recent surprising one is that of coupling-constant metamorphosis \cite{Hietarinta1984coupling}, whereas under suitable conditions in a classical Hamiltonian system it is possible to promote a coupling constant to the role of a new dual Hamiltonian, while the original Hamiltonian becomes a coupling constant itself.  This allowed the discovery of a series of dualities between systems previously considered different, as for example the Henon-Heiles and Holt systems. A sizeable literature has arisen around the subject and we refer the interested reader to related articles. 
 
Other types of dualities  have been studied in the seemingly different context of the Eisenhart-Duval null lift of a natural Hamiltonian system: this is a higher dimensional description of a quadratic Hamiltonian system given in terms of null geodesics \cite{Eisenhart1928,DuvalBurdetKunzlePerrin1985,GaryDuvalHorvathy1991,Minguzzi2006,Minguzzi2007}. Such dualities have been employed to map the standard Kepler problem to a Dirac-type theory of gravity with time dependent gravitational constant \cite{GaryDuvalHorvathy1991}, as well as to  relate the motion of a particle in a electric and magnetic field to that into a new set of generally time-dependent fields \cite{LyndenBellWonder,LyndenBellWandering,cariglia2014conformal}, or to discuss transformations that are related to the appearence of a cosmological constant like term \cite{gibbons2014dark}. In all these cases, the dual systems are such that their Eisenhart-Duval metrics are related by a change of variables plus a conformal rescaling. The Eisenhart-Duval lift has found a renewed interest in recent years due to its application to the non-relativistic AdS/CFT duality,  see for example \cite{duval2009geometry,son2008toward,balasubramanian2008gravity,bekaert2013embedding} and the related vast bibliography. An important ingredient in the construction is the fact that the Eisenhart-Duval lift provides a Bargmann structure, which in turn defines a Newton-Cartan structure which has the interpretation of a non-relativistic spacetime on which it is possible to describe dynamics, including gravity. As an application the Schr\"{o}dinger-Newton equation has been generalised  in \cite{ChristianLazzarini2015} using the Eisenhart-Duval lift to include a curved spatial background and non-inertial forces, and its Schr\"{o}dinger-Newton maximal group of symmetries has been calculated. Both \cite{gibbons2014dark} and \cite{ChristianLazzarini2015} discussed the role of the Schwarzian derivative in the conformal rescaling of the metric: the first noticed how it is related to a cosmological-constant type term, and the second in a related fashion required a condition of zero Schwarzian derivative of the (integral of the) conformal factor of the transformation in order to preserve the source term of the gravitational potential for the Schr\"{o}dinger-Newton equation. 
 
In this work we present a unifying point of view on the subjects above, in terms of what we call projective geometry of the dynamics. A quadratic Hamiltonian system admits null lifts which are not unique, of which the Eisenhart-Duval one is a specific example with special properties. To a null lift one can associate a section in a bundle of projective conics, which are defined as quadratic forms on a projective tangent space. The resulting dynamics is a theory of unparameterised geodesic curves, and the same section can be projected to seemingly different Hamiltonian system, thus clarifying that the related dualities are just alternative descriptions of the same dynamics. The operation of changing the parameter on the geodesics induces a conformal rescaling of the lift metric, and since the underlying structure is independent of the choice of parameter a Weyl geometry is naturally induced. M\"{o}bius transformations play a  special role since they are in a way the mildest possible kind of conformal rescalings, as they leave invariant the trace-free Ricci tensor. We describe in detail the dualities that act on an Eisenhart-Duval null lift, and provide a number of examples as well as a projective description of the coupling-constant metamorphosis and the Jacobi metric. We also describe the basic elements of the quantum version of the theory. 
 
The rest of the work is organised as follows. In sec.\ref{sec:preliminary} we describe basic notions of the main objects we will deal with: null lifts, conformal rescalings and changes of parameterisation, Weyl geometry, the Schwarzian tensor. The following sec.\ref{sec:general} contains the main theory: the projective tangent bundle, projective conics and Hamiltonian dualities. We also describe how conformal Killing tensors are natural objects in a Weyl structure. Sec.\ref{sec:examples} deals with examples and applications of the general theory. We provide a new projective interpretation of the Jacobi metric and the coupling-constant metamorphosis, as well as a number of examples that show how previously known dualities fit into the description of the same projective object. Sec.\ref{sec:quantum} describes the quantum case applied to Eisenhart-Duval lifts, where projective dualities carry over from the classical to the quantum theory using the Yamabe operator. We finish in sec.\ref{sec:conclusions} with a summary and conclusions.

\section{Preliminary notions\label{sec:preliminary}} 
\subsection{Natural Hamiltonians and null lifts\label{sec:null_lifts}} 
A natural Hamiltonian is by definition given by a quadratic function of the momenta 
\be \label{eq:natural_Hamiltonian}
H = \frac{1}{2} h^{ij} (p_i + e A_i ) (p_j + e A_j) + e^2 V  \, , 
\ee 
where $\{q^i, p_i \}$, $i=1, \dots, \dm$ are conjugate variables, $e$ a constant, $h^{ij}(q,t)$ an inverse metric, $V(q,t)$ a scalar potential, $A_i(q,t)$ a vector potential and the variable $t$ is time. A null lift of this system corresponds to a Hamiltonian $\cal{H}$ that is: i) non-degenerate  , non-definite, quadratic and homogeneous in a new set of momenta that includes the original ones, ii) such that the extra momenta are all conserved, and iii) such that the original Hamiltonian can be recovered by setting $\cH =0$. An important example is given by the Eisenhart-Duval null lift\cite{Eisenhart1928,DuvalBurdetKunzlePerrin1985,GaryDuvalHorvathy1991} 
\be 
\cH = \frac{1}{2} h^{ij} (\h{p}_i - \h{p}_v A_i ) (\h{p}_j - \h{p}_v A_j) + \h{p}_v^2 V + \h{p}_u \h{p}_v \, , 
\ee 
where new conjugate variables $(u, \h{p}_u)$, $(v, \h{p}_v)$ have been introduced, and with a slight abuse of notation by $h$, $V$ and $A_i$ we indicate a new set of fields that depend on $q^i$ and $u$, and are related to the quantities in the original Hamiltonian via \eqref{eq:t_u}. The conserved quantity $\h{p}_v$ is associated to translations along $v$ which are symmetries of $\cH$: then one can perform as Marsden-Weinstein reduction eliminating both $v$ and $\h{p}_v$. The choice $\h{p}_v = 0$ is allowed but not of interest in the present context, as it leads to a lower dimensional Hamiltonian with no scalar and vector potential. Setting $\h{p}_v = e \neq 0$ and $\cH = 0$ instead one obtains the condition $e \h{p}_u = - H$, where as will be seen right below $\h{p}_i = - p_i$. Then the new variables can be projected out yielding the original system. The condition on $\h{p}_u$ can be re-written as 
\be \label{eq:t_u}
t = - \frac{u}{e} \, , 
\ee 
where $t$ is a new time variable that we can introduce as associated to the Hamiltonian flow of the original Hamiltonian $H$. Let us call $\lambda$ the evolution parameter for the trajectories of $\cH$. Then the equation of motion for $u$ is $\frac{d u}{d \lambda} = \frac{\partial \cH}{\partial \h{p}_u} = \h{p}_v = e$, which implies that $t = - \lambda$ modulo a constant term. This minus sign is of historical rather than fundamental origin, and is responsible for the fact that $\h{p}_i = - p_i$, see \cite{cariglia2014conformal} for details. 
 
Writing $\cH = \frac{1}{2} g^{AB} p_A p_B$, with $A, B = 1, \dots, \dm +2$, $p_A = (\h{p}_i, \h{p}_u, \h{p}_v)$, from the Eisenhart-Duval lift we can extract the Lorentzian metric 
\be \label{eq:Eisenhart_metric}
ds^2 = g_{AB} \, dy^A dy^B = h_{ij} dq^i dq^j + 2 du \left( dv - V du + A_i dq^i \right) \, , 
\ee 
with $y^A = (q^i, u, v)$. This metric belongs to a family of metrics with degenerate curvature invariants \cite{hervik2014degenerate}. 
 
The null lift of a Hamiltonian is not unique. The following is an example of different null lift : 
\be 
\cH =  \frac{1}{2} h^{ij} (p_i + p_v A_i ) (p_j + p_v A_j) + p_v^2 V \pm p_T^2 \, .  
\ee 
This will provide an invertible metric if $V$ is never zero. Given that the potential is defined modulo a constant it will be sufficient  that $V$ admits a global minimum. Then the Hamiltonian above reduces to that of the original system upon choosing $p_v = e$, $p_T^2 = \mp H$, where the choice of sign used depends on the sign of the energy $H$. For this reason  this kind of lift can be used to describe either positive or negative energy states at a time, but not both. 
 
A second example of null lift can be given for the case where the Hamiltonian is of the form 
\be 
H = \frac{1}{2} h^{ij} p_i p_j  + e_1^2 V_1 + e_2^2 V_2 \, , 
\ee 
then one can choose 
\be 
\cH = \frac{1}{2} h^{ij} \h{p}_i \h{p}_j + \h{p}_{v_1}^2 V_1 + \h{p}_{u_1} \h{p}_{v_1} + \h{p}_{v_2}^2 V_2 + \h{p}_{u_2} \h{p}_{v_2}\, . 
\ee 
In other words, one can lift separately the two potentials and the reduction if performed by imposing $\h{p}_{v_1} = e_1$, $\h{p}_{v_2} = e_2$, $e_1 \h{p}_{u_1} + e_2 \h{p}_{u_2} = - H$. This gives a metric that is always non-degenerate, and the same can be done for more than two pairs of $\{ u, v \}$ variables. Metrics of this kind, and generalizations including vector potential-type terms, have been studied in \cite{hervik2014degenerate}. Another type of reduction of metrics of this kind in which instead one takes $\h{p}_{u} = 0$ appears in \cite{GaryMarco2013}, where each coupling constant of the Toda chain is independently lifted. We conclude this section by noticing that quadratic Hamiltonians are not necessarily only associated to non-relativistic dynamics, for example a relativsitc particle in a scalar potential is also described by a quadratic Hamiltonian.

\subsection{Conformal rescaling as change of parameterisation\label{sec:conformal_rescalings}} 
Let $g_{AB}(y)$ be an indefinite metric, with associated geodesic Hamiltonian 
\be 
\cH = \frac{1}{2} g^{AB}(y) p_A p_B \, . 
\ee 
Null geodesics solve Hamilton's equations
\ba 
&& \dl{y^A} = \{ y^A, \cH \} = g^{AB} p_B \, , \\ 
&& \dl{p_B} = \{ p_B, \cH \} = - \frac{1}{2} \partial_B (g^{RS}) p_R p_S \, ,  
\ea 
where $\{ \cdot, \cdot \}$ is the Poisson bracket, 
plus the condition $\cH=0$.

Consider a new Hamiltonian associated to the rescaled tensor $\bb{g}_{AB} = \Omega^2 (y) g_{AB}$: 
\be \label{eq:tilde_H} 
\bb{\cH} = \frac{1}{2} \Omega^{-2} (y) g^{AB} \, p_A p_B \, , 
\ee 
where the factor $\Omega$ is almost everywhere different from zero. 
It is a well known fact that $\bb{\cH}$ and $\cH$ share the same null geodesics: in fact the new equations of motion are for $\mathcal{H} = 0$: 
\ba 
\dlb{y^A} &=& \Omega^{-2} g^{AB} p_B + \frac{\partial \, \Omega^{-2}}{\partial p_A} \cH  \nn \\ 
&=& \Omega^{-2}  g^{AB} p_B = \Omega^{-2}  \dl{y^A} \, , \\ 
\dlb{p_B} &=& - \frac{\Omega^{-2}}{2} \partial_B (g^{RS}) p_R p_S -  \partial_B \left( \Omega^{-2}\right) \cH \nn \\ 
&=& - \frac{\Omega^{-2}}{2} \partial_B (g^{RS}) p_R p_S = \Omega^{-2} \dl{p_B}\, . 
\ea 
In fact on each null geodesic at a time we can change the evolution parameter according to  
\be 
d\bb{\lambda} = \Omega^{2}(y(\lambda)) \, d\lambda \, , 
\ee 
and this maps the equations of motion for $\bb{\cH}$ into those for $\cH$. The transformation just described is
a Weyl rescaling of the metric, for which $\bb{\cH}$ is a geodesic Hamiltonian as well. Importantly, we can
consider a more general rescaling of the type $\Omega = \Omega (y, p)$ : this in general will map to a non-geodesic $\bb{\cH}$, but it can happen that a canonical transformation makes it geodesic again. In the rest of this work, with the exception of sec.\ref{sec:Jacobi}, we limit ourselves to the purely geometric case $\Omega = \Omega(y)$. However it should be noted that a dependence on both the $y$ and $p$ variables in the general case corresponds to a much greater amount of freedom. 
 
In this work we will deal with two types of 'conformal transformations'. The first  type is a pure \textit{Weyl rescaling} of the metric $g \rightarrow \Omega^2 g$: as discussed in this section, such a rescaling corresponds to a change of parameterisations of null geodesics. The second type is a change of metric induced by a map $f$ of the manifold in itself, such that the pullback of a metric $\bb{g}$ under $f$ is  conformally related to another metric $g$: $f^* \bb{g} = \Omega^2 g$. This type of conformal transformation we call \textit{conformal diffeomorphism}, see sec.\ref{sec:classification}. While in the classical theory both types are relevant, in the quantum theory there are no trajectories to which one can associate a change of parameter, and the relevant type of transformation that seems to be relevant is only conformal diffeomorphisms, see sec.\ref{sec:quantum}.

\subsection{Nuts and bolts of Weyl geometry\label{sec:Weyl}} 
A Weyl structure is obtained specifying the triple $\{ g_{AB}, \nabla , \varphi_A \}$, where $g_{AB}$ is a metric, $\nabla$ a covariant derivative $\nabla$ and $\varphi_A$ a field with the following properties. 
 
$\nabla$ is torsionless and non-metric, with a specific type of non-metricity. By non-metricity we mean that 
\be 
\nabla_A g_{BC} = \partial_A g_{BC} - \Gamma_{AB}^L g_{LC} - \Gamma_{AC}^L g_{BL} \neq 0 \, ,  
\ee 
where $\Gamma_{AB}^C$ are the Christoffel symbols. In the specific case of a Weyl structure the non-metricity is such that 
\be \label{eq:non_metricity_Weyl} 
\nabla_A g_{BC} = 2 \varphi_A g_{BC} \, , 
\ee 
where $\varphi_A$ is a vector field that quantifies the fact that parallel transport does not preserve the length of vectors. In fact if we transport a vector $V^A$ along a curve $\lambda \mapsto \gamma(\lambda)$ with tangent vector $T(\lambda)$ according to $T^A \nabla_A V^B = 0$, then from \eqref{eq:non_metricity_Weyl} one obtains the variation of $V^2 = g_{AB} V^A V^B$: 
\be 
\frac{d V^2}{d \lambda} = T^A \nabla_A V^2 = 2 \,  T^A \varphi_A V^2 \, . 
\ee 
Using the property of absence of torsion eq.\eqref{eq:non_metricity_Weyl} can be used to solve for the Christoffel symbols obtaining 
\be \label{eq:Weyl_Christoffel}
\Gamma_{AB}^C = {}_g\Gamma_{AB}^C + g_{AB} \varphi^C - \delta_A^C \varphi_B - \delta_B^C \varphi_A \, , 
\ee 
where ${}_g\Gamma_{AB}^C$ are the Levi-Civita Christoffel symbols. Finally, as last ingredient of a Weyl structure one specifies a transformation law for $g$ and $\varphi_A$  under a change of scale: 
\ba 
g_{AB} &\rightarrow& \bb{g}_{AB} = \Omega^2 (y) g_{AB} \, , \label{eq:change_of_scale_g} \\ 
\varphi_A &\rightarrow& \bb{\varphi}_A = \varphi_A + \partial_A \ln \Omega \, , \label{eq:change_of_scale_varphi}
\ea 
and takes the quotient under the equivalence relationship defined by the change of scale. Since under\eqref{eq:change_of_scale_g} the Levi-Civita Christoffel symbols change by 
\be \label{eq:Christoffel_change}
{}_{\bb{g}}\bb{\Gamma}^A_{BC} = {}_g \Gamma^A_{BC} + \Omega^{-1} \left( \partial_B \Omega \delta^A_C + \partial_C \Omega \delta^A_B - \partial^A \Omega g_{BC} \right) \, ,  
\ee 
then using \eqref{eq:change_of_scale_varphi} the definition \eqref{eq:Weyl_Christoffel} yields invariant Christoffel symbols 
\be 
\bb{\Gamma}_{AB}^C = \Gamma_{AB}^C \, . 
\ee

Any function $F$, whether a scalar or tensor, that transforms as 
\be 
F \rightarrow \bb{F} = \Omega^w F 
\ee 
is called a \textit{Weyl field} of weight $w$. The metric for example has weight 2. The Weyl covariant derivative $\nabla = \partial + \Gamma$ is not scale covariant when acting on fields on non-zero weight. So one can define a scale-covariant derivative 
\be \label{eq:D1}
D F = \nabla F - w \varphi \otimes F \, , 
\ee 
which satisfies 
\be \label{eq:D2} 
\bb{D} \bb{F} = \Omega^w D F \, . 
\ee

 Different interpretations have been given historically to the change of scale above according to the different application of Weyl geometry, in our case we will consider it as a local change of units.

\subsection{Conformal transformations and the Schwarzian tensor\label{sec:preliminaries_schwarzian}}  
The Schwarzian derivative has recently appeared in a number of dualities between natural Hamiltonian systems \cite{gibbons2014dark,cariglia2014conformal} and in the description of symmetries of the Schr\"{o}dinger-Newton equation \cite{ChristianLazzarini2015}. It is also naturally related to conformal and projective transformations. In this subsection we draw upon \cite{gibbons2014dark,osgood1992schwarzian} and describe the Schwarzian tensor that generalises the concept of Schwarzian derivative to dimension greater than two. Let $\cM$ be a manifold of dimension $N$ with metric $g$ and $\varphi$ a real function on $\cM$. 
Let $\mathcal{M}$ be another manifold with the same dimension and another metric 
$\bb{g}$ and $f: \cM \rightarrow \bar{\mathcal{M}}$ be a conformal diffeomorphism with 
pullback of $g$ given by $f^* \bb{g} = e^{2\varphi} g$. Then the Schwarzian tensor of $\varphi$ with respect to $g$ is defined by 
\ba 
{}_g B_{MN} (\varphi) &=& {}_g\nabla_M \, {}_g\nabla_N \varphi - {}_g \nabla_M \varphi \, {}_g\nabla_N  \varphi \nn \\ 
&& - \frac{1}{N} \left( {}_g\nabla^2 \varphi - \left({}_g\nabla \varphi \right)^2 \right) g_{MN} \, , 
\ea 
and it is used to define the Schwarzian derivative of $f$ by $(Sf)_{MN} = {}_g B_{MN}(\varphi)$. It also makes sense to talk about the Schwarzian derivative of a conformal rescaling. Under a composition of conformal transformations 
\be 
\{ \cM, g \} \overset{f, e^{2\varphi}}{\xrightarrow{\hspace*{1cm}} } \{ \bb{\cM}, \bb{g} \} \overset{g, e^{2\sigma}}{\xrightarrow{\hspace*{1cm}} } \{ \bb{\bb{\cM}}, \bb{\bb{g}} \}
\ee 
 there is the useful property  
\be 
{}_g B_{MN} (\varphi + \sigma ) = {}_g B_{MN}(\varphi) + {}_{\bb{g}} B_{MN}(\sigma) \, , 
\ee 
or 
\be 
S(g \circ f) = S(f) + f^* S(g) \, . 
\ee 
A \textit{M\"{o}bius transformation} is a conformal rescaling or a conformal diffeomorphism such that ${}_g B(\varphi) = 0$. 
 
One of the main properties of the Schwarzian tensor is that under a conformal rescaling or a conformal diffeomorphism it expresses the change in the trace-free Ricci tensor 
\be 
R^{(0)}_{MN} = R_{MN} - \frac{1}{N} R g_{MN} \, , 
\ee 
as 
\be 
R^{(0)}_{MN} = f^* \bar{R}^{(0)}_{MN} + (N - 2) {}_g B_{MN}(\varphi) \, . 
\ee 
Einstein spaces are those for which $R^{(0)}_{MN} = 0$. From the property above it is clear that a characterisation of M\"{o}bius transformations is that the leave unchanged the trace-free Ricci tensor. 
 
In the specific case of a $u$-dependent conformal rescaling of an Eisenhart-Duval metric, $f^* \bb{g} = \varphi^\prime (u) g$, in \cite{ChristianLazzarini2015} it has been shown that the Schwarzian tensor is given by 
\be \label{eq:Schwarzian_1}
{}_g B =  \frac{1}{2} S(\varphi) du \otimes du \, , 
\ee 
where $S(\varphi)$ is the standard Schwarzian derivative 
\be \label{eq:Schwarzian_2}
S(\varphi) = \frac{\varphi^{\prime\prime\prime}}{\varphi^\prime} - \frac{3}{2} \left(\frac{\varphi^{\prime\prime}}{\varphi^\prime} \right)^2 \, . 
\ee

\section{The general theory\label{sec:general}} 
\subsection{Projective tangent bundle} 
In this work for simplicity we deal with a phase space $\mathcal{P}$ that is the cotangent bundle of a base manifold $\mathcal{M}$, $\cal{P} = T^* \cal{M}$. Let $N$ be the dimension of $\cal{M}$, $a \in \cM$ represent a point on $\cM$ and let $T \mathcal{M}$ be the tangent bundle. We want to endow $T\cM$ with a projective structure, or in other words we want to be able to identify, on each tangent space $T_{a} \cM$ vectors under the equivalence relation 
\be \label{eq:er_vectors}
V^\mu \sim \pm \omega^2 V^\mu \, , \quad V^\mu \neq 0 \, , 
\ee 
where $\omega \in \mathbb{R}^*$ is a constant. In other words we are interested in undirected line segments. The projective structure on each tangent space $T_{a} \cM$ smoothly joins from point to point on the base manifold $\cM$, yielding a smooth bundle structure called \textit{projective tangent bundle of $\cM$}, denoted with the symbol $\ptm$. This can be achieved for example in the following way. The standard construction of each tangent space $T_{a} \cM$ is done by considering all curves $\lambda \mapsto \gamma(\lambda)$ that pass through $a$ at $\lambda = 0$, $\lambda(0) = a$, and then defining the tangent vector $V = \left.\frac{d\gamma}{d\lambda}\right|_{\lambda=0}$ by the action $V(f) := \left.\frac{d f}{d\lambda}\right|_{\lambda=0}$ $\forall f \in C^{\infty} \cM$. We can inherit the equivalence relation \eqref{eq:er_vectors} if we identify curves $\gamma$ modulo a change of parameterisation. We will say locally that $\frac{d\gamma}{d \lambda} \sim \frac{d\bar{\gamma}}{d{\lambda}}$ if $\gamma(0) = a = \bar{\gamma}(0)$ and if there exists a locally invertible $C^\infty$ function $\lambda \mapsto \bar{\lambda}(\lambda)$ such that $\bb{\gamma}(\lambda) = \gamma(\bb{\lambda}(\lambda))$ in a neighbourhood of zero. Then 
\be 
\frac{d\bb{\gamma}}{d\lambda} = \left. \frac{d\bb{\lambda}}{d\lambda}\right|_{\lambda = 0} \, \frac{d\gamma}{d{\lambda}} \, . 
\ee 
The requirement that the change of parameterisation is locally invertible assures that $\left. \frac{d\bb{\lambda}}{d\lambda}\right|_0 \neq 0$, and we can identify $\frac{d\bb{\lambda}}{d\lambda}$ with $\pm \omega^2$ in \eqref{eq:er_vectors}. One can move from local to global by considering a theory of unparameterised curves on $\cM$. Doing so, and asking the change of parameterisation is always invertible, means that the plus or minus sign in \eqref{eq:er_vectors} can be chosen only once along the whole curve, however one retains the freedom to fix pointwise the value of $\frac{d\bb{\lambda}}{d\lambda}$ arbitrarily. So from the start our theory will be a theory of unparameterised curves on $\cM$. The structure group $GL(\dm, \mathbb{R})$ of $T\cM$ naturally induces a structure group $\mathbb{P} GL(\dm, \mathbb{R})$ of projective transformations. 
 
\subsection{Null dynamics as a choice of projective conics} 
Now let's introduce dynamics through a null geodesic, i.e. homogeneous second order in momenta, null Hamiltonian $\mathcal{H}$. The chief example we have in mind for this work is when $\mathcal{H}$ is the null lift of a natural Hamiltonian as in  \eqref{eq:natural_Hamiltonian}, however in this section we will just assume a null geodesic Hamiltonian is given, regardless of its origin. In the case of a null lift we will assume that coordinate space is $n$ dimensional  and that $N = n + m$, where $N$ is the dimension of the lift; for example in the case of the Eisenhart-Duval lift $m=2$, while for the last example in section \ref{sec:null_lifts} $m=4$. 
Let $\mathcal{H}$ be given in terms of a metric as 
\be 
\mathcal{H} = \frac{1}{2} g^{AB}(y) p_A p_B \, ,  
\ee 
where $y^A$ are coordinates for the lift manifold $\mathcal{M}$. 
Then Hamilton equations imply $\frac{dy^A}{d\lambda} = g^{AB} p_B$, therefore we can write the null condition as 
\be \label{eq:projective_conic}
 \frac{1}{2} g_{AB}(y) \frac{dy^A}{d\lambda} \frac{dy^B}{d\lambda} = 0 \, . 
\ee 
This is the equation of a projective conic on $\ptm$, so we can think of the choice of $\mathcal{H}$ as a point by point choice of projective conics, or a section in a bundle of projective conics with base $\cM$. Because of the projective nature of the conic, a different choice of metric $\bb{g} = \Omega^2 (y) g$ induces the same conic, and therefore what is relevant is the conformal class of $[g]$ and an exact Weyl geometry is naturally induced: one can start with a reference metric $g \in [g]$ and its Levi-Civita covariant derivative, which implies $\varphi = 0$ in this gauge, and then transform $g$ and $\varphi$ according to \eqref{eq:change_of_scale_g}, \eqref{eq:change_of_scale_varphi}. By exact geometry we mean that in every gauge $\varphi_A dy^A$ is an exact form. If one chooses  a different metric $\bb{g} = \Omega^2 (y) g$ from the conformal class and defines a new null Hamiltonian $\bb{\mathcal{H}}$ with it, then by the results of section \ref{sec:conformal_rescalings} we know that $\bb{\mathcal{H}}$ and $\mathcal{H}$ share the same unparameterised null geodesics, so the choice of $\mathcal{H}$ is compatible with the projective structure. Hence, assuming sufficent regularity in the coefficients, we have proved the following: \\ 
\textit{Projective conics on $\ptm$ are in a one to one correspondence with a null dynamics described by unparameterised curves}. 

The dynamical trajectories then can be locally described by $N - 2$ functions as follows. First   one uses \eqref{eq:projective_conic} to isolate one of the velocities as a function of the others, for example writing $\frac{dy^N}{d\lambda} = f(y, \frac{dy^{\hat{A}}}{d\lambda})$, $\hat{A} = 1, \dots, N-1$. Next, assuming that for example at a point $\frac{dy^{N-1}}{d\lambda} \neq 0$ then locally one can invert $\lambda = \lambda (y^{N - 1})$ and express the remaining functions as $y^i (y^{N - 1})$, $i = 1, \dots, N -2$. 
 
We conclude this section with a remark. When $\mathcal{H}$ is the null lift of a natural Hamiltonian $H$, then one can think of \eqref{eq:projective_conic} as originating from a  \textit{projective completion} of an affine variety on the cotangent bundle $T^*\cM$ that is built naturally from $H$. For example, in the Eisenhart-Duval lift $\mathcal{H}$ is a projective completion of the affine variety defined by 
\be 
H - E = 0 \, , 
\ee 
defined on $T^* \mathcal{M}$, where $E$ corresponds to the energy. The completion in general is not unique.

\subsection{Classification of projective conics and Hamiltonian dualities\label{sec:classification}}

We now consider diffeomorphisms of the lift manifold $\mathcal{M}$: $f: \cM \rightarrow \cM$, $y \mapsto \bb{y}$. In coordinates  
\ba 
\bar{y}^A &=& \bar{y}^A (y^B) \, , \label{eq:diff1} \\ 
\frac{d\bb{y}^A}{d\lambda} &=& \frac{\partial \bb{y}^A}{\partial y^B} \frac{dy^B}{d\lambda} \, . \label{eq:diff2}
\ea 
The metric $g$ is related to $\bb{g}$ by pullback: $g = f^* \bb{g}$, or in terms of matrices    
\be \label{eq:projective_transformation}
g = J^T \bar{g} \, J  \, , 
\ee 
where $J^A {}_B = \frac{\partial \bar{y}^A}{\partial y^B} \in GL(\dm, \mathbb{R})$. This  induces a natural action of $\mathbb{P} GL(\dm, \mathbb{R})$ on the conformal class $[\bb{g}]$, which we will denote by $f^* [\bb{g}]$, and with the same reasoning from eq.\eqref{eq:diff2} one see that the pullback acts on $\ptm$. 
 
We recall from standard projective geometry that the pullback, acting on symmetric bilinear forms, yields a left action of $GL(\dm , \mathbb{R})$ that preserves the number of null and positive definite spaces of $g$. The action is not transitive and we can consider its orbits. A fundamental result is that any two bilinear forms $\bar{g}$, $g$ on $T\cM $ 
that have the same dimension of null and positive definite spaces under this action are on the same orbit. \footnote{For projective conics there is a further $\mathbb{Z}_2$ freedom to change the bilinear form from $g$ to $-g$, which in our case corresponds to changing the sign of $\mathcal{H}$.} We will use this fact to relate Hamiltonian systems one to another. By this we mean the following. Suppose $\mathcal{H}$ is the null lift Hamiltonian of a dynamical system, with associated conformal class of metrics $[g]$. Now, if along the orbit of $[g]$ under the projective transformations \eqref{eq:projective_transformation} one can find a different equivalence class $[g^\prime]$ that is also the null lift of another Hamiltonian system, we will say that the two lower dimensional systems are \textit{projectively equivalent}. In particular, one can divide projective equivalence into two main broad cases: 
\begin{itemize} 
\item $[g^\prime] = f^*[\bb{g}]  = [\bb{g}]$, the same conformal class: this is a \textit{dynamical symmetry} of the lower dimensional system into itself; 
\item $[g^\prime] = f^*[\bb{g}] \neq [\bb{g}]$: this is a \textit{projective duality} between the two systems. 
\end{itemize}

Pointwise the theory is uninteresting since for example in the concrete case of the Eisenhart-Duval lift for any two different natural Hamiltonian systems with the same number of degrees of freedom their lifts define equivalent projective conics and can be mapped one into the other by a $\mathbb{P} GL(\dm , \mathbb{R})$ transformation. In this setting this is the same as the statement that any pseudo-Riemannian metric can be put in Lorentzian form at a single point by a coordinate transformation. 
The theory becomes instead interesting when considering a local or global situation. In this case we can say that two different Hamiltonian systems with the same number of degrees of freedom are projectively equivalent if the associated sections in the bundle of projective conics are projectively equivalent. By saying this we mean that there exists a transformation as in \eqref{eq:diff1}, \eqref{eq:diff2} such that at all points one projective conic is mapped to the other, or equivalently $\bar{g}$ is mapped into $g$ modulo a positive factor $\Omega^2 (y)$. 
 
As a concrete example we take as null lift of a natural Hamiltonian system its Eisenhart-Duval lift.  If $\bb{g}$ is of the Eisenhart-Duval type \eqref{eq:Eisenhart_metric}, then $ f^* \bb{g} = \Omega^2 g$ will have $g$ also of Eisenhart-Duval type if $g_{vv} = 0 = g_{iv}$. This can be written as 
\ba 
\bb{g}_{AB} J^A_{(v)} J^{B}_{(v)} &=& 0 \, , \label{eq:constraint_projective1} \\ 
\bb{g}_{AB} J^A_{(v)} J^{B}_{(i)} &=& 0 \, . \label{eq:constraint_projective2} 
\ea 
Then the factor $\Omega^2$, the metric $h_{ij}$ and the potentials $V$ and $A_i$ can be obtained from 
\ba 
\Omega^2 &=& \bb{g}_{AB} J^A_{(u)} J^{B}_{(v)} \, , \\ 
\Omega^2 h_{ij} &=& \bb{g}_{AB} J^A_{(i)} J^{B}_{(j)} \, , \\ 
- 2 \Omega^2 V &=& \bb{g}_{AB} J^A_{(u)} J^{B}_{(u)} \, , \\ 
\Omega^2 A_i &=& \bb{g}_{AB} J^A_{(u)} J^{B}_{(i)} \, . 
\ea 
These equations represent the most general conformal transformation that preserves the form of an Eisenhart-Duval metric, and hence its most general projective duality. 
 
It is useful to specialise these transformations to transformations that preserve the conformal Bargmann structure. In \cite{GaryDuvalHorvathy1991,ChristianLazzarini2015} it was shown these are given by transformations such that 
\be \label{eq:Bargmann_preserving} 
f^*\bb{g} = \Omega^2 (u) g \, , \quad f_* \partial_v = \frac{1}{\nu} \partial_v \, , 
\ee 
where $\nu \in \mathbb{R}^*$ necessarily. The latter condition is equivalent to 
\be \label{eq:Bargmann_preserving_condition1}
\frac{\partial \bb{y}^i }{\partial y^v } = 0 \, , \quad \frac{\partial \bb{y}^u }{\partial y^v } = 0 \, , \quad \frac{\partial \bb{y}^v }{\partial y^v } = \frac{1}{\nu} \, . 
\ee 
Then  \eqref{eq:constraint_projective1} is automatically satisfied, while \eqref{eq:constraint_projective2} implies 
\be \label{eq:Bargmann_preserving_condition2} 
J^u_{(i)}=0 \, . 
\ee 
The remaining projective duality conditions are given by 
\ba 
\Omega^2 &=& \frac{1}{\nu} J^u_{(u)}  \, , \label{eq:Bargmann_preserving_phi}  \\ 
\Omega^2 h_{ij} &=& \bb{h}_{lm} J^l_{(i)} J^{m}_{(j)} \, , \\ 
- 2 \Omega^2 V &=& \bb{g}_{AB} J^A_{(u)} J^{B}_{(u)} \, , \\ 
\Omega^2 A_i &=& \bb{h}_{lm} J^l_{(u)} J^{m}_{(i)} + \bb{A}_l J^l_{(i)} J^u_{(u)} \nn \\ 
&& +  J^u_{(u)} J^v_{(i)} \, . 
\ea 
We can further specialise to \textit{temporal re-parameterisations}, that are transformations that, in addition to the properties above, satisfy 
\be \label{eq:temporal_q}
\bb{q}^i = \rho (u) q^i \, , 
\ee 
i.e. the $q$ coordinates transform with a time dependent rescaling. This gives $J^i_{(j)} = \rho \, \delta^i_j$, $J^i_{(u)} = \rho^\prime q^i$. For these the projective duality conditions become 
\ba 
\Omega^2 &=& \frac{1}{\nu} J^u_{(u)}  \, ,  \label{eq:temporal_phi} \\ 
\Omega^2 h_{ij} &=& \rho^2 \bb{h}_{ij}  \, ,  \label{eq:temporal_h}  \\ 
- 2 \Omega^2 V &=& (\rho^\prime)^2 \bb{h}_{lm} q^l q^m + 2 \nu \Omega^2 \rho^\prime \bb{A}_l q^l \nn  \\ 
&& - 2 \nu^2 \Omega^4 \bb{V} + 2 \nu \Omega^2 J^v_{(u)}  \, , \label{eq:temporal_V} \\ 
\Omega^2 A_i &=& \rho \rho^\prime \bb{h}_{il} q^l + \nu \Omega^2 \rho \bb{A}_i   + \nu \Omega^2 J^v_{(i)} \, . \label{eq:temporal_A}
\ea

\subsection{Conformal Killing tensors and conserved quantities\label{eq:conformal_Killing_tensors}} 
Conserved quantities for null geodesic motion that are polynomial in momenta are constructed from conformal Killing tensors. Conformal Killing tensors keep attracting attention, see for example \cite{casalbuoni2014conformal} where they are studied in the context of conformal dynamics of relativistic particles. 
A conformal Killing tensor of rank $p$ is a symmetric tensor $K_{M_1 \dots M_p} = K_{(M_1 \dots M_p)}$, where the round brackets mean symmetrisation of indices, that satisfies the differential equation 
\be \label{eq:conformal_Killing}
{}_g \nabla_{(M} K_{N_1 \dots N_p )} = g_{(M N_1} T_{N_2 \dots N_p)} \, , 
\ee 
where the rank $(p-1)$ tensor $T$ is related to the divergence of $K$ and the derivatives of its traces, see \cite{MarcoRMP} for examples and further references. From a conformal Killing tensor it is possible to build the following conserved quantity 
\be 
C = K^{M_1 \dots M_p} p_{M_1} \dots p_{M_p} \, . 
\ee 
A conformal Killing tensor is natural object for a Weyl geometry as we now show. 
Under a rescaling of the metric $g \rightarrow \bar{g} = \Omega^2 g$ the Christoffel symbols change according to \eqref{eq:Christoffel_change}. 
As a consequence, the symmetrised covariant derivative of $K$ becomes 
\ba 
{}_{\bb{g}}\bb{\nabla}_{(M} K_{N_1 \dots N_p )} &=& {}_g \nabla_{(M} K_{N_1 \dots N_p )} \nn \\ 
&& \hspace{-3cm} - 2p \Omega^{-1} \left( \partial_{(M} \Omega K_{N_1 \dots N_p)} - \frac{1}{2} g_{(M N_1} \partial^R \Omega K_{R N_2 \dots N_p)} \right) \nn \, . 
\ea
Then, it follows that $\Omega^{2p} K_{M_1 \dots M_p}$ is a conformal Killing tensor for the metric $\bb{g}$, from which one can build the conserved quantity 
\ba 
\bb{C} &=& \bb{g}^{M_1 N_1} \dots \bb{g}^{M_p N_p} \left( \Omega^{2p} K_{N_1 \dots N_p} \right) \bb{p}_{M_1} \dots \bb{p}_{M_p} = C \nn \, ,   
\ea 
where we used $\bb{p}_A = p_A$. 
To summarise, under a conformal rescaling of the metric the tensor $K_{M_1 \dots M_p}$ with lower indices is a Weyl field of weight $2p$, and the tensor with upper indices $K^{M_1 \dots M_p}$ is invariant. The conserved quantity $C$ is both a scalar and invariant under rescalings. 
In the case we are considering we take $g$ to be the fiducial metric for which $\varphi_A = 0$, so that ${}_g \nabla = D$ in this gauge. Then we can rephrase the result in the language of the scale covariant derivative defined in \eqref{eq:D1}, \eqref{eq:D2}: since $T$ in \eqref{eq:conformal_Killing} has weight $2p-2$ we can write the manifestly scale invariant equation 
\be 
D_{(M} K_{N_1 \dots N_p )} = g_{(M N_1} T_{N_2 \dots N_p)} \, . 
\ee

In the next section we discuss a number of examples and applications of the general theory, noting the role of M\"{o}bius transformations when relevant. The first examples describe  temporal re-parameterisations obtained using the Eisenhart-Duval lift. Then we describe more non-trivial projective dualities as the one leading to the Jacobi metric and to the coupling-constant metamorphosis.  \\ 
 
\section{Applications and examples\label{sec:examples}} 
 
\subsection{Jacobi metric\label{sec:Jacobi}} 
This is the only section where we consider a momentum dependent rescaling of the Hamiltonian that will lead to the well known Jacobi metric. The starting point is a classical mechanical system: 
\be 
H = \frac{1}{2} h^{ij}(q) p_i p_j + V(q) \, . 
\ee 
We separately lift positive and negative energy trajectories using the null Hamiltonian 
\be 
\cH =  \frac{1}{2} h^{ij}(q) p_i p_j + V(q) p_y^2 - \sgn (H) p_z^2 \, . 
\ee 
Choosing $p_y^2 = 1$, $p_z^2 = |E|$, where $E$ is the energy associated to $H$, one gets null geodesics that project down to the original system. Now we rescale the Hamiltonian times the factor $\Omega^{-2} (q,p) = (\sgn (H) p_z^2 - V(q) p_y^2)^{-1}$: 
\be 
\bb{\cH} =  \frac{1}{2} \frac{h^{ij}(q) p_i p_j}{\sgn (H) p_z^2 - V(q) p_y^2} - 1  \, . 
\ee 
Notice that this is no longer a homogeneous quadratic Hamiltonian. On null geodesics with $p_y^2 = 1$ and $p_z^2 = |E|$ this reduces to 
\be 
\bb{\cH} =  \frac{1}{2} \frac{h^{ij}(q) p_i p_j}{E - V(q)} - 1 \, , 
\ee
which is the null geodesic Hamiltonian related to the Jacobi-metric. The relationship between the $T$ time variable of $\bar{\cH}$ and the original $t$ variable of $\cH$ is 
\be 
\frac{dT}{dt} = \sgn (H) p_z^2 - V p_y^2  \, ,  
\ee 
which reduces to the familiar reparameterisation $\frac{dT}{dt} = E - V$. Finally, it is worth noticing that $\bar{\cH} = 0$ is equivalent to say the curve is parameterised by the arc-length. 

\subsection{Coupling-constant metamorphosis} 
Coupling-constant metamorphosis is a type of duality between Hamiltonian systems that allows to exchange the role of the Hamiltonian with that of a coupling-constant. It was first discussed in \cite{Hietarinta1984coupling} and has given rise to a fairly wide bibliography. We refer the reader to articles citing the original one mentioned above. As by no meaning exhaustive examples of further results we mention \cite{kalnins2010coupling}, where higher order symmetries in the momenta where discussed, and \cite{sergyeyev2012coupling} where the duality was extended to general finite-dimensional dynamical systems. 
 
In \cite{Hietarinta1984coupling} one starts from a Hamiltonian 
\be \label{eq:ccm1}
H = H_0 (q,p) - g F(q) \, , 
\ee 
that is integrable for any value of the coupling constant $g$ and switches to a different Hamiltonian 
\be \label{eq:ccm2}
G = \frac{H_0}{F} - \frac{h}{F} \, . 
\ee 
Formally, $G$ is obtained by isolating the $g$ term in \eqref{eq:ccm1} and promoting it to a phase-space function, while at the same time changing the Hamiltonian $H$ into a constant $h$. Then $G$ is integrable for any value of $h$ and if $\{ I_j (q,p,g) \}$ are integrals of motion of $H$, then $J_j = I_j (q,p, h)$ are integrals of motion for $G$. It is also shown that the time derivative under the Hamiltonian $H$ with time $t$ and the Hamiltonian $G$ with time $T$ are related by 
\be \label{eq:Hietarinta_time_change} 
dT = F dt \, . 
\ee 
We now discuss these results in the context of natural Hamiltonians and projective geometry. 
Suppose $H_0 = \frac{1}{2} h^{ij} p_i p_j + V(q)$ is a natural Hamiltonian, then we can consider the following null lift 
\be 
\cH = \frac{1}{2} h^{ij} p_i p_j + V(q) p_y^2 - F \, p_w^2 - \sgn(H^\prime) p_z^2 \, , 
\ee 
where $H^\prime = \frac{1}{2} h^{ij} p_i p_j + V(q) p_y^2 - F \, p_w^2$. 
Choosing $p_y^2 = 1$, $p_w^2 = g$, and $p_z^2 = |h|$ projects back to the original system. We are assuming  here $g > 0$, which is not restrictive modulo a change of sign for $F$. We rescale the Hamiltonian by a factor of $\Omega^{-2} = F^{-1}$: 
\ba 
\bar{\cH} &=& \frac{\cH}{F} = \frac{1}{2 F} h^{ij} p_i p_j + \frac{V(q)}{F} p_y^2 -  p_w^2 - \frac{ \sgn(H^\prime) p_z^2}{F} \nn \\ 
&=& \frac{ \left( \frac{1}{2 } h^{ij} p_i p_j + \frac{V(q)}{F} p_y^2 \right) - \sgn(H^\prime) p_z^2}{F} - p_w^2 
\ea 
obtaining a null lift of the dual Hamiltonian \eqref{eq:ccm2}. The conformal rescaling corresponds to a change in time parameter that is the same as \eqref{eq:Hietarinta_time_change}. 
 
For what concerns conserved quantities, both $\cH$ and $\bar{\cH}$ are homogeneous and of degree two in the momenta, and therefore if we restrict ourselves to discussing polynomial in momenta conserved quantities, then these are associated to conformal Killing tensors. As seen in section \ref{eq:conformal_Killing_tensors}, we can choose the same tensor with upper indices for both metrics. Then the same quantity 
\be 
C = K^{A_1 \dots A_p} p_{A_1} \dots p_{A_p} = C(p_i , p_y, p_w, p_z) 
\ee 
will project to $C (p_i , g, |h|)$ in one system, and $C(p_i, G, |h|)$ in the other.

\subsection{Temporal re-parameterisations for a scalar potential} 
Temporal re-parameterisations in a theory with a scalar potential have been considered in \cite{gibbons2014dark}, where it has been shown that in general a cosmological constant-like term arises in the scalar potential after the transformation. We remind here the reader of the main result, adapting the notation to the one used in this work. Given a theory with a scalar potential $\bb{U}$ and Eisenhart-Duval metric 
\be 
\bb{g} = d\bb{q}^i d\bb{q}^i + 2 d\bb{u} d\bb{v} - 2 \bb{V} d\bb{u} d\bb{v} \, , 
\ee 
then the following change of variables 
\be \label{eq:dark_energy_change_of_variables}
\bb{u} = \varphi (u) \, , \quad \bb{q}^i  = \sqrt{|\varphi^\prime|} q^i \, , \quad \bb{v} = \sgn \varphi^\prime \left( v - \frac{\varphi^{\prime\prime}}{\varphi^\prime} \frac{q^i q^i}{4} \right) \, , 
\ee 
is such that the pullback of $\bb{g}$ is given by 
\be 
g = f^* \bb{g} = |\varphi^\prime | \left( dq^i dq^i + 2 du dv - 2 V du^2 \right) \, , 
\ee 
with 
\be \label{eq:dark_energy_V}
V (q^i, u) = |\varphi^\prime| \bb{V} (  \sqrt{|\varphi^\prime|} q^i, \varphi(u) ) + \frac{1}{4}S(\varphi) q^i q^i \, . 
\ee 
The second term plays the role of a cosmological constant in Newtonian cosmology if $S(\varphi) = const.$

To make contact with the results of the previous section, we note that from the change in the metric it must be that $|\varphi^\prime| = \Omega^2$, while from \eqref{eq:dark_energy_change_of_variables} and \eqref{eq:temporal_q} one reads  $\rho =  \sqrt{|\varphi^\prime|}$. Then $\Omega^2 = \rho^2$, which implies in \eqref{eq:temporal_h} that $h = \bb{h}$ as seen above. Eq.\eqref{eq:dark_energy_change_of_variables} also implies that $\nu = \sgn \varphi^\prime$, and then this agrees with \eqref{eq:temporal_phi} which reads $|\varphi^\prime| = \sgn \varphi^\prime \frac{\partial \bb{u}}{\partial u}$. Then, an explicit check of eqs.\eqref{eq:temporal_V}, \eqref{eq:temporal_A} shows that $A_i = 0$ and that $V$ is given by \eqref{eq:dark_energy_V}. 
 
We note from the results of sec.\ref{sec:preliminaries_schwarzian} that M\"{o}bius transformations are those for which no cosmological constant-type term is generated.

\subsection{Transformations of electric and magnetic fields} 
In \cite{cariglia2014conformal} similar transformations\footnote{For notational purposes, one should note that in \cite{gibbons2014dark,cariglia2014conformal} the current function $\varphi$, written there as $f$, is expressed as a function of the variable $t = - u$. This explains some sign changes.} with a curved spatial metric $h_{ij}$ 
\be \label{eq:LB_change_of_variables}
\bb{u} = \varphi (u) \, , \quad \bb{q}^i = \sqrt{|\varphi^\prime|} q^i \, , \quad \bb{v} = \sgn \varphi^\prime \left( v - \frac{1}{4} \frac{\varphi^{\prime\prime}}{\varphi^\prime} h_{ij} q^i q^j \right) \, , 
\ee 
in the specific case of $S(\varphi)=0$, 
change the metric  
\be 
\bb{g} = \bb{h}_{ij} d\bb{q}^i d\bb{q}^j + 2 d\bb{u} (d\bb{v} - \bb{V} d\bb{u} +  \bb{A}_i d\bb{q}^i ) \, , \ee 
which in this case describes a scalar and vector electric potential into 
\be 
g = f^* \bb{g} = |\varphi^\prime | \left( h_{ij} dq^i dq^j + 2 du ( dv - 2 V du^2 + A_i dq^i ) \right) \, , 
\ee 
where $\bb{h}_{ij} = h_{ij}$ and  the potentials are related by 
\ba 
V &=& |\varphi^\prime| \bb{V} - \frac{\varphi^{\prime\prime}}{2 |\varphi^\prime|} \bb{q}^i \bb{A}_i \, , \\ 
A_i &=& \sgn \varphi^\prime \sqrt{|\varphi^\prime|} \bb{A}_i \, . 
\ea 
One can explicitly check again that these changes of potentials are the same as those given by eqs.\eqref{eq:temporal_V} and \eqref{eq:temporal_A}.

\subsection{Dirac-type theory of gravity} 
In \cite{GaryDuvalHorvathy1991} solutions of Newton's equations of gravity for a time dependent potential 
\be 
\bb{V} = - G(\bb{u}) \frac{M}{\bb{q}} \, , 
\ee 
were mapped into those of a standard time independent potential 
\be 
V = - G_0 \frac{M}{q} \, , 
\ee 
by the following transformations 
\be 
\bb{q}^i = \Omega(u) q^i \, , \quad \bb{u} = - \frac{a^2}{u+b} + c \, , \quad \bb{v} = v + \frac{q^2}{2(u+b)} + d \, , 
\ee 
where 
\be 
\Omega (u) = \frac{a}{u+b} \, , \quad G_0 = G(\bb{u}) | \Omega (\bb{u}) | \, . 
\ee 
The Eisenhart-Duval metric 
\be 
\bb{g} = d\bb{q}^i d\bb{q}^i + 2 d\bb{u} d\bb{v} - 2 \bb{V} d\bb{u} d\bb{v} 
\ee 
transforms under pullback into 
\be 
g = f^* \bb{g} = \Omega^2(u) \left( dq^i dq^i + 2 du dv - 2 V du^2 \right) \, , 
\ee  
We recognise here that $\rho = \Omega$, which implies $\bb{h}_{ij} = h_{ij}$. Also $\nu = 1$, which is consistent with \eqref{eq:temporal_phi} since $\frac{\partial \bb{u}}{\partial u} = \Omega^2$.  The changes of potentials are the same as those given by eqs.\eqref{eq:temporal_V} and \eqref{eq:temporal_A}. 
 
We note that the conformal factor in the pullback can be written as $\Omega^2 = \varphi^\prime$, with $\varphi = - \frac{a^2}{u+b}$, which is a fractional linear function, that in other words satisfies $S(\varphi) = 0$. Then this transformation is M\"{o}bius. \\

\subsection{The Schr\"{o}dinger group of transformations} 
Given a flat $\dm+2$ dimensional Eisenhart-Duval metric, i.e. which lifts a free natural Hamiltonian in $\dm$ dimensions with no potentials, the Schr\"{o}dinger group $\text{Sch}(\mathbb{R}^{\dm+1,1})$ is by definition the group of conformal diffeomorphisms of the lift spacetime that keep invariant the conformal Bargmann structure, or in other words such that \eqref{eq:Bargmann_preserving}  holds with $\nu = 1$ \cite{GaryDuvalHorvathy1991}. Part of these symmetries survive in the presence of monopoles \cite{horvathy1983dynamical,horvathy2006dynamical}. 
For a flat Eisenhart-Duval metric the full transformations are given for example in \cite{ChristianLazzarini2015}: 
\ba 
\vec{\bb{q}} &=&  \frac{A \vec{q} + \vec{b} u + \vec{c}}{f u + g} \, , \label{eq:chrono1} \\ 
\bb{u} &=& \frac{d u + e}{f u + g} \, , \label{eq:chrono2} \\ 
\bb{v} &=&  v + \frac{f}{2} \frac{ | A \vec{q} + \vec{b} u + \vec{c} |^2}{f u + g} - \vec{b} \cdot A \vec{q}  \nn \\ 
&&  \hspace{3.2cm} - \frac{|\vec{b}|^2 u}{2} + h  \, , \label{eq:chrono3}
\ea 
where $A \in O(\dm)$ is related to rotations, $\vec{b}$ to boosts, $\vec{c}$ to translations, the matrix 
\be 
D = \left( \begin{array}{cc} d & e \\ f & g \end{array} \right) \in GL(2, \mathbb{R}) 
\ee 
is associated to the projective group of the time axis, and $h \in \mathbb{R}$ is a central extension parameter. For these transformations the factor $\Omega^2$ is given by 
\be 
\Omega^2 = \frac{1}{(f u + g)^2} \, , 
\ee 
and $ dg - ef = 1$. In particular, $\Omega^2 = \varphi^\prime$, with $\varphi = \frac{(d u + e)}{ (f u + g)}$, and therefore these are all M\"{o}bius transformations. 
 
It has been known for some time that the centerless Schr\"{o}dinger group is isomorphic to a group of matrices of the kind 
\be \label{eq:matrix_projective}
\left( \begin{array}{ccc} A^i_j & b_j & c_j \\ 0 & d & e \\ 0 & f & g \end{array} \right) 
\ee 
that acts projectively on the $(q^i, u)$ coordinates as 
\ba 
\left( \begin{array}{ccc} A^i_j & b^j & c^j \\ 0 & d & e \\ 0 & f & g \end{array} \right)  
	\left( \begin{array}{c} q^i \\ u \\ 1 \end{array} \right) &=& 
	\left( \begin{array}{c}(Aq + b u + c)^i  \\ du + e \\ fu + g \end{array} \right) \nn \\ 
&\sim& \left( \begin{array}{c}\frac{(Aq + b u + c)^i}{fu + g}  \\ \frac{du + e}{fu + g} \\ 1 \end{array} \right) \, , 
\ea 
see for example \cite{DuvalThesis1982,burdet1983chronoprojective}. We want here to give a projective interpretation of the result. Suppose a transformation is given that changes a flat Eisenhart-Duval lift into another one, modulo a conformal factor $\Omega^2$, i.e. 
\be \label{eq:chronoprojective_pullback}
f^* (\bb{g}) = f^* \left( d\bb{q}^i d\bb{q}^i + 2 d\bb{u} d\bb{v} \right) = \Omega^2 \left( dq^i dq^i + 2 du dv \right) = \Omega^2 g \, , 
\ee 
and that this transformation preserves the Bargmann structure as in \eqref{eq:Bargmann_preserving}. Then the following diagram holds 
\be 
g = \Omega^{-2} f^* (\bb{g}) \xleftarrow{\bb{\lambda} \rightarrow \lambda} f^* (\bb{g}) = \Omega^2 g \xleftarrow{f} \bb{g} \, ,  
\ee 
meaning that one can start with a geodesic of $\bb{g}$, satisfying $\frac{d^2 \bb{y}^A}{d \bb{\lambda}^2} = 0$, then apply $f^{-1}$ to obtain a curve $y^A (\bb{\lambda })$ that is a geodesic of the metric $\Omega^2 g$, and then using the results of section \ref{sec:conformal_rescalings} to define a  new parameter $\lambda (\bb{\lambda})$ such that 
\be \label{eq:chrono_parameter_change}
\frac{d\lambda}{d \bb{\lambda}} = \Omega^{-2} \, . 
\ee 
Then, the curve $y^A (\lambda)$ is also a geodesic, satisfying $\frac{d^2 y^A}{d\lambda^2} = 0$, if we initially started with a null geodesic of $\bb{g}$. Writing explicitly the equations of motion for the middle metric with 'mixed' curves $y^a (\bb{\lambda})$ one finds after basic manipulations and using \eqref{eq:chrono_parameter_change} that \footnote{One can explicitly check that using these equations  to calculate $\frac{d^2 \bb{y}^A}{d \bb{\lambda}^2} = 0$ from the full transformation \eqref{eq:chrono1}-\eqref{eq:chrono3} using \eqref{eq:chrono_parameter_change} agreement is found for null geodesics.}
\ba 
\frac{d^2 q^i}{d{\lambda}^2} &=& 0 \, , \label{eq:chrono_mixed_1} \\ 
\frac{d^2 u}{d{\lambda}^2} &=& 0 \, , \label{eq:chrono_mixed_2} \\ 
\frac{d^2 v}{d{\lambda}^2} &=& \frac{\partial_u \Omega}{\Omega} \left( \frac{dq^i}{d{\lambda}} \frac{dq^i}{d{\lambda}} + 2 \frac{du}{d\lambda} \frac{dv}{d\lambda} \right) \label{eq:chrono_mixed_3} \, . 
\ea 
The importance of eqs.\eqref{eq:chrono_mixed_1}-\eqref{eq:chrono_mixed_3} is that they show that applying $f^{-1}$ to the $\bb{y}$ variables induces a transformation on the $(q^i, u)$ variables that maps lines into lines, modulo a rescaling factor. When $\Omega^2 = 1$ then these are exactly affine transformations, and for $\Omega^2 \neq 1$ they generalise affine transformations. In both cases they are projective transformations and can be obtained projectively using the matrix in eq.\eqref{eq:matrix_projective}, where we used the information that conditions \eqref{eq:Bargmann_preserving_condition1}, \eqref{eq:Bargmann_preserving_condition2} must hold. Then it follows automatically that $\bb{u} (u)$ is a M\"{o}bius transformation. The transformation rule for $\bb{v} = \bb{v} (q^i, u, v)$ is not independent and can be obtained from the transformation rule of the metric, which gives the full Jacobian of the change of variables, and the known $(q^i, u)$ transformations.

\section{An outline of the quantum case\label{sec:quantum}} 
The discussion so far has been valid for an arbitrary type of null lift, however it is difficult to discuss the quantum case without specialising to specific types of lifts. We will discuss here mainly the case of the Eisenhart-Duval lift for which the classical dualities can be carried on to the quantum version of the theory. 
 
The important object in the quantum theory is the Yamabe operator, or conformal Laplacian \cite{DuvalOvsienko2001,Besse1987,michel2014second}. For a given metric $g$ its conformal laplacian or Yamabe operator is given by 
\be 
\Delta_Y = \nabla_A g^{AB} \nabla_B - \frac{N-2}{4(N-1)} R \, , 
\ee 
where the first term is the ordinary Laplacial and $R$ is the scalar curvature of $g$. $\Delta_Y$ is called conformal since under a rescaling of the metric $g \rightarrow \bb{g} = \Omega^2 g$ it transforms as 
\ba \label{eq:Yamabe_rescaling}
\bb{\Delta}_Y &=& \Omega^{- \frac{N-2}{2}} \Delta_Y \circ \Omega^{\frac{N-2}{2}} \, . 
\ea 
It is well known that for an Eisenhart-Duval lift metric $g$ as in \eqref{eq:Eisenhart_metric} the standard Laplacian induces the quantum Schr\"{o}dinger equation \cite{GaryDuvalHorvathy1991}: the equation $\Delta \Psi = 0$ is equivalent to 
\ba \label{eq:Schrodinger}
i \hbar \partial_t \psi = - \frac{\hbar^2}{2m} \frac{1}{\sqrt{h}} D_i \left( \sqrt{h} h^{ij} D_j \right) + U \psi \, , 
\ea 
where $D_i$ is the gauge covariant derivative and using a wavefunction $\Psi = e^{- \frac{i m}{\hbar}} \psi(q,u)$, plus $u = - t$, $V = \frac{U}{m}$. 
 
In the quantum theory the Yamabe operator can be used to carry over projective dualities at the price of including the Ricci scalar term in the scalar potential. This can be viewed as a quantum correction to the classical duality. In other words we replace the equation $\Delta \Psi = 0$ with $\Delta_Y \Psi = 0$ \cite{ChristianLazzarini2015}, which induces a Schr\"{o}dinger equation in the same form as \eqref{eq:Schrodinger}, with the difference that the potential term is replaced by 
\be 
U^\prime = U + \frac{\hbar^2}{m}  \frac{N-2}{8(N-1)} R \, . 
\ee 
Now, suppose a projective duality relates the metric $g$ and $\bb{g}$ with $f^* \bb{g} = \Omega^2 g$ as in sec.\ref{sec:classification}, and that both metrics are Eisenhart-Duval lifts. Since the conformal Laplacian is covariant it gives the same result whether it is calculated using the $\bb{y}$ or $y$ coordinates, and therefore $\bb{\Delta}_Y$ and $\Delta_Y$ are related  by eq.\eqref{eq:Yamabe_rescaling}. Considering the wavefunction $\bb{\Psi} = \Omega^{- \frac{N-2}{2}} \Psi$ then $\bb{\Delta}_Y \bb{\Psi} = 0$ if and only if $\Delta_Y \Psi = 0$, and therefore one Schr\"{o}dinger equation transforms into the other. 
 
Lastly, we mention the interesting fact that in \cite{michel2014second} quantum symmetry operators of order 1 and 2 in derivatives are explicitly calculated, as well as their relative quantum anomalies. This is relevant to generalise the classical polynomial conserved quantities and to study R-separability of variables. We do not plan to discuss the quantum case further in this work, the main aim of this section being discussing the projective duality.

\section{Conclusions\label{sec:conclusions}} 
In this work we have described quadratic Hamiltonian systems in terms of projective geometry and Weyl structures, thus providing a unifying point of view on their dualities. It would be interesting to know if this new point of view can be fruitfully applied to novel problems. One  question that comes to mind is if one can give a projective interpretation of the phenomenon of R-separation of the Schr\"{o}dinger equation \cite{KalninsMiller1982intrinsic,KalninsMiller1984,michel2015second}. We also notice how the results discussed here imply that recent efforts in obtaining new non-trivial Einstein spaces from dynamical systems \cite{cariglia2015ricci,filyukov2015self} can be extended to the more general case of Einstein-Weyl spaces, see \cite{calderbank1997einstein}. A considerable amount of research has been produced in the area, as a non-exhaustive example see \cite{kiosak2009there,kuhnel2009einstein,Brinkmann1925einstein}. We leave this to future work.

\vspace{0.2cm}

\section*{Acknowledgments}
The author would like to thank C. M. Warnick, A. Galajinsky and the anonymous reviewer for reading the manuscript and useful comments, in particularly the latter for thoroughly reading the work. The author acknowledges funding by PROPP, Federal University of Ouro Preto.  


\newpage  
\bibliographystyle{apsrmp}
\bibliography{Bibliography_06Apr2015}
%
%
%
%


\end{document}